\documentclass[conference]{IEEEtran}
\usepackage{amsfonts}
\IEEEoverridecommandlockouts

\usepackage{amsmath,amssymb}
\usepackage{array}
\usepackage{booktabs}       
\usepackage{stfloats}      
\usepackage{graphicx}
\usepackage{subcaption}

\usepackage{algorithm}
\usepackage{algorithmic}
\ifCLASSINFOpdf
\else
\fi

\usepackage{cite}
\usepackage{float}
\usepackage{color}
\begin{document}
\title{Do Ambient Backscatter Communication Receivers Require Low-Noise Amplifiers?}\vspace{-25pt}
\author{Xinyi Wang, Yuxin Li, Yinghui Ye, Gongpu Wang, and Guangyue Lu \\ \vspace{-25pt}
\thanks{Xinyi Wang, Yuxin Li, Yinghui Ye, and Guangyue Lu are with  Xi'an University of Posts \& Telecommunications, Xi'an 710121, China. }
\thanks{Gongpu Wang is with Beijing Jiaotong University, Beijing 100044, China. 

This work was supported by the National Natural Science Foundation of China under Grants
62571430, and 62471388, and the Innovation Capability Support Program of Shaanxi under Grant 2024ZC-KJXX-016.
}}
\maketitle
\begin{abstract}
In ambient backscatter communication (AmBC), strong direct interference from the ambient source poses a major challenge to reliable symbol detection. Although previous studies have shown that employing a low-noise amplifier (LNA) in conventional point-to-point communication improves symbol detection performance at low-to-moderate transmission power, it remains unclear whether this improvement also holds for AmBC. To respond it, in this work, we investigate the symbol detection performance of an AmBC receiver that is equipped with an LNA and adopts the energy detection (ED) to recover tag's information. Particularly, we first propose a new AmBC symbol detection framework that incorporates LNA parameters.  On this basis, we derive the bit error rate (BER) of the ED and employ the deflection coefficient (DC) to demonstrate that the detection performance can be enhanced by the LNA at low-to-moderate ambient source transmission power. Then, we derive the near-optimal detection threshold to minimize the BER and propose a method to estimate the required parameters for this threshold by leveraging the tag's pilot symbols.
\end{abstract}
\section{Introduction}
Ambient backscatter communication (AmBC) is an emerging green wireless communication technology with great potential for Internet of Things (IoT) applications \cite{10463656}. In AmBC, the tag transmits its information by modulating the incident ambient signal, rather than generating its own carrier wave, thereby enabling low-power transmission. Consequently, the AmBC receiver simultaneously receives both the direct ambient source signal and the backscattered signal, leading to a major challenging in symbol detection.

The authors in \cite{8007328} proposed the energy detection (ED) with a near-optimal detection threshold, which is estimated by using the information symbols transmitted by the tag, to minimize bit error rate (BER). The authors in \cite{10531099} and \cite{10220163} proposed an improved ED by replacing the squaring operation of the received sample in the ED with an arbitrary positive power operation. In \cite{9250656}, a deep transfer learning-based symbol detection framework was developed, where convolutional neural networks were utilized to extract features from the covariance matrix of the received sample. The coding techniques have also been adopted at the tag to enhance the detection performance. In \cite{7551180} and \cite{10005249}, a differential encoding technique was considered and an energy difference based detection was proposed, and a closed-form symbol detection threshold was derived to minimize BER. Other studies like using the non-return-to-zero coding, and the orthogonal space-time block coding can also be found in \cite{9242274}, and \cite{9430725}, respectively.

In existing research \cite{8007328,10531099,10220163,9250656,7551180,10005249,9242274,9430725}, it was assumed that the received radio frequency (RF) signal is directly  down converted to baseband and sampled for ED without using low-noise amplifier (LNA). {\color{black}However, in traditional point-to-point communication systems, it has been well-established that in regions of low-to-moderate transmission power, employing an LNA can improve symbol detection performance. This observation motivates us to explore the potential benefits of integrating an LNA into the AmBC receiver. }However, there are two key differences between traditional point-to-point communication and AmBC that introduce uncertainty about whether or not the use of an LNA in AmBC receiver improves symbol detection performance. First, in the traditional  point-to-point communication, both the useful signal and antenna noise at the receiver are amplified, whereas in AmBC, the tag's signal (useful signal), the strong direct interference and the antenna noise at both the receiver and the tag are amplified. Second, due to the different signal components at the receiver, the corresponding  distortion signals caused by the non-linear behavior of an LNA under both systems are different.

 To answer this question, we systematically model an AmBC receiver architecture equipped with an LNA, and analyze the BER of the ED. The main contributions are summarized below.
\begin{itemize}
\item We propose a  symbol detection framework for AmBC that incorporates LNA parameters. Considering an on-off keying (OOK) modulation scheme at the tag and the ED, we derive a closed-form expression for the BER and the near-optimal detection threshold to minimize  BER.
\item {\color{black}We conduct an analysis based on the deflection coefficient (DC) to demonstrate that the integration of an LNA boosts the difference in the probability density function (PDF)  of the energy of the samples under \textquotedblleft 0\textquotedblright\ and \textquotedblleft 1\textquotedblright\ transmitted by the tag at low-to-medium transmission power.}
\item The near-optimal threshold requires prior knowledge of parameters such as the channel coefficient, LNA, noise power, and transmit power of the ambient source, which are typically unknown in practical systems. To address this, we propose to leveraging the statistical properties of pilot signal samples to estimate the necessary parameters for calculating detection threshold.
\end{itemize}

 The main notations  are listed below. $\mathbb{E}\left[ x \right]$, ${\mathop{\rm var}} \left[ x \right]$, and $\left| x \right|$ denote the expectation, the variance, and the absolute value of $x$, respectively. $Q\left( x \right) = \frac{1}{{\sqrt {2\pi } }}\int_x^\infty  {\exp \left( { - \frac{{{t^2}}}{2}} \right)} dt$ is the Q function. $\mathbb{CN}\left( {a,b} \right)$  is the Gaussian distribution with mean $a$ and variance $b$.
\section{System Model}
\begin{figure}
  \centering
  \includegraphics[width=0.3\textwidth]{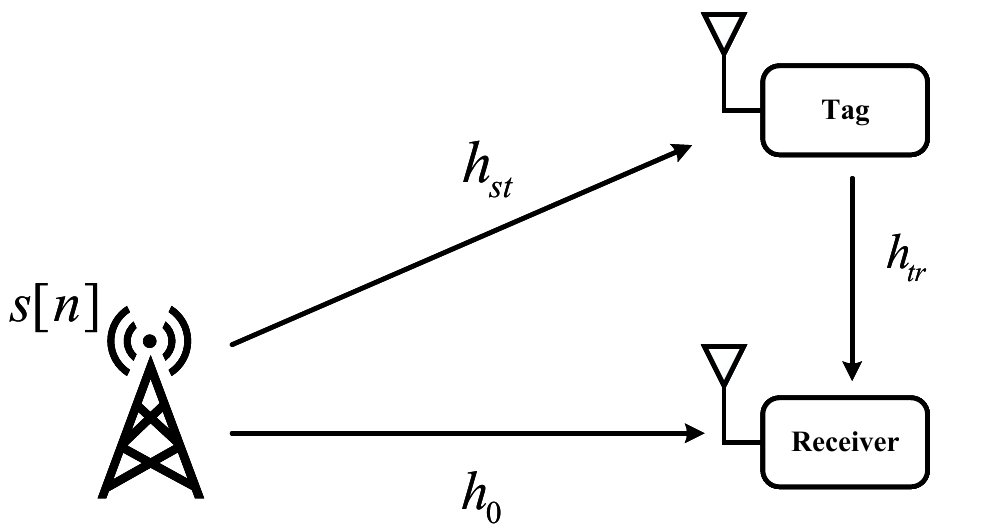}\\
  \caption{A three-node ambient backscatter system.}\label{fig1}\vspace{-10pt}
\end{figure}

The system model under consideration is shown in Fig. \ref{fig1}. The transmitted RF signals are modulated and reflected by the tag via OOK. All channels undergo independent quasi-static fading. Let ${h_0}$, ${h_{st}}$ and ${h_{tr}}$ represent the complex channel coefficients of the ambient source-receiver link, the ambient source-tag link, and the tag-receiver link, respectively. We assume that the tag transmits $K$ symbols in each coherent time interval. 
\subsubsection{Signal Model Without LNA}
\begin{figure}
  \centering
  \includegraphics[width=0.42\textwidth]{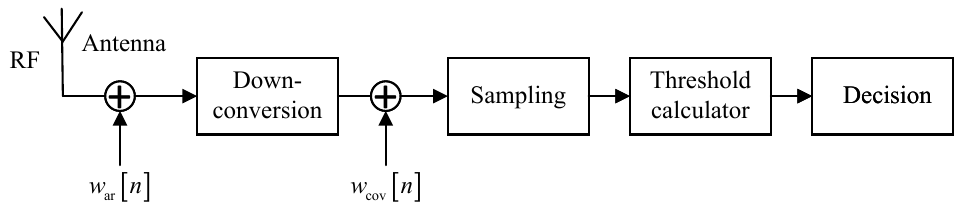}\\
  \caption{Architecture of a conventional AmBC receiver.}\label{fig2}
\end{figure}

Fig. \ref{fig2} illustrates the architecture of a conventional AmBC receiver. The RF signal is first captured by the antenna, a process which introduces antenna noise ${w_{\rm{ar}}}\left[ n \right] \sim \mathbb{CN}\left( {0,{N_{\rm{ar}}}} \right)$. Subsequently, the signal is down-converted to baseband, which introduces down-conversion noise ${w_{{\mathop{\rm cov}} }}\left[ n \right] \sim \mathbb{CN}\left( {0,{N_{{\mathop{\rm cov}} }}} \right)$.

Since the tag's symbol rate is much lower than that of the ambient source, we assume that the tag's symbol remains constant over $N$ consecutive samples $s\left[ n \right]$ of the ambient source.
{\color{black}Then, the $n$-th received sample of the $k$-th symbol for the conventional receiver is written as}{\footnote{\color{black}{Here, we assume that the receiver employs an ideal analog-to-digital converter (ADC) with infinite resolution and operates away from clipping, in order to isolate and quantify the individual effects of the LNA. Introducing ADC quantization noise into the system model only requires updates to the received signal power and noise components. Thus, our analysis framework and detection methods remain applicable with ADC quantization noise.}}}
\begin{align}
{y_k}\left[ n \right] =
\underbrace{\alpha {h_{st}}{h_{tr}}d\left[ k \right]s\left[ n \right]}_{\text{useful signal}} +
\underbrace{{h_0}s\left[ n \right]}_{\text{direct interference}} +
\underbrace{w\left[ n \right]}_{\text{noise}},
\end{align}
where $n = 1, \ldots ,N$, $k = 1, \ldots ,K$, $d\left[ k \right] \in \left\{ {0,1} \right\}$ denotes the tag's symbols, $s\left[ n \right] \sim \mathbb{CN}\left( {0,{P_s}} \right)$, $w\left[ n \right] = {w_{{\rm{ar}}}}\left[ n \right] + {w_{{\rm{cov}}}}\left[ n \right] + \alpha {h_{tr}}d\left[ k \right]{w_{{\rm{at}}}}\left[ n \right]$, ${w_{\rm{at}}}\left[ n \right] \sim \mathbb{CN}\left( {0,{N_{\rm{at}}}} \right)$ represents the antenna noise at the tag, $w\left[ n \right] \sim \mathbb{CN}\left( {0,{N_w}} \right)$, ${N_w} = {N_{{\rm{ar}}}} + {N_{{\rm{cov}}}} + {\alpha ^2}{\left| {{h_{tr}}} \right|^2}{N_{{\rm{at}}}}d\left[ k \right]$, and $\alpha $ is the tag coefficient representing the scattering efficiency and antenna gain.

\subsubsection{Signal Model with LNA}

\begin{figure}
  \centering
  \includegraphics[width=0.46\textwidth]{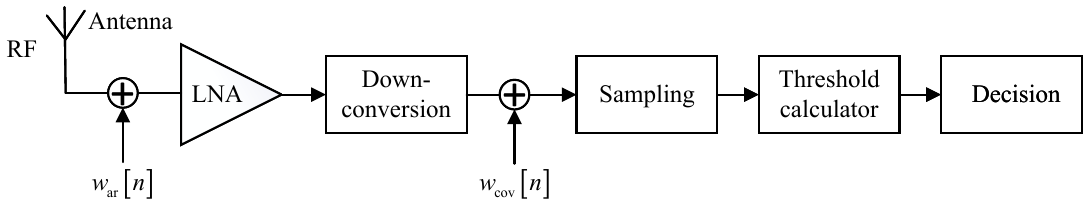}\\
  \caption{Architecture of the AmBC receiver  with an LNA.}\label{fig3}\vspace{-4pt}
\end{figure}

The architecture of the AmBC receiver equipped with an LNA is shown in Fig. \ref{fig3}. Clearly, the LNA  amplifies not only the useful signal but also the direct interference and noise components ${w_{\rm{at}}}\left[ n \right]$ and ${w_{\rm{ar}}}\left[ n \right]$. Additionally, the intermodulation distortion (IMD) is introduced by the LNA due to its inherent nonlinearity. In this work, only odd-order nonlinear terms are considered in the modeling of the RF front-end nonlinearity, since the spurious components generated by even-order terms-such as harmonics and intermodulation/crossmodulation products-are assumed to fall outside the narrowband spectrum of interest\cite{haghghadam2011contributions}. Moreover, the IMD of the fifth order and above can be neglected as its power is sufficiently lower compared to the third order. Under these assumptions, the baseband equivalent signal at the receiver can be expressed as (\ref{2}), as shown at the top of next page,
\begin{figure*}[t]  
\centering
\begin{align}\label{2}
y_k^{\mathrm{a}}[n] &= \underbrace{\beta_1\alpha h_{st}h_{tr}d[k]s[n]}_{\text{useful signal}}
                     + \underbrace{\beta_1 h_0 s[n]}_{\text{direct interference}}
                     + \underbrace{\beta_3(h_0[n] + \alpha h_{st}h_{tr}d[k])s[n]|(h_0[n] + \alpha h_{st}h_{tr}d[k])s[n]|^2}_{\text{IMD}}
                     + \underbrace{w_{\mathrm{a}}[n]}_{\text{noise}} \nonumber \\
&= \begin{cases}
\beta_1 h_0 s[n] + \beta_3 h_0 s[n] |h_0 s[n]|^2 + w_{\mathrm{a}}[n],\!\!&{\mathcal{H}_0} \\
\beta_1 h_1 s[n] + \beta_3 h_1 s[n] |h_1 s[n]|^2 + w_{\mathrm{a}}[n],\!\!&{\mathcal{H}_1}
\end{cases}
\end{align}
\hrulefill
\end{figure*}
where ${h_1} = {h_0} + \alpha {h_{st}}{h_{tr}}$, ${\mathcal{H}_i}$ denotes the hypothesis with $d\left[ k \right] = i$, ${w_{\rm{a}}}\left[ n \right] = {\beta _1}{w_{{\rm{ar}}}}\left[ n \right] + {w_{{\rm{cov}}}}\left[ n \right] + {\beta _1}\alpha {h_{tr}}d\left[ k \right]{w_{{\rm{at}}}}\left[ n \right]$, ${w_{\rm{a}}}\left[ n \right] \sim \mathbb{CN}\left( {0,N_{\rm{aw}}} \right)$, ${N_{{\rm{aw}}}} = {\beta _1}^2{N_{{\rm{ar}}}} + {N_{{\rm{cov}}}} + {\beta _1}^2{\alpha ^2}{\left| {{h_{tr}}} \right|^2}{N_{{\rm{at}}}}d\left[ k \right]$, and ${\beta _1}$, ${\beta _3}$ are characteristics of the receiver front-end. As reported in Section II of \cite{4939441}, the typical values of $\beta_1$ and $\beta_3$ are ${\rm{56}}{\rm{.23}}$, and ${\rm{ - 7497}}{\rm{.33}}$, respectively.

 \section{Performance of an ED with LNA}

 By adopting ED, the tag's symbol can be detected by using the following rule, given by \cite{8007328}
\begin{align}
\begin{array}{l}
\hat d\left[ k \right] = 0,\left\{ {\begin{array}{*{20}{l}}
{\begin{array}{*{20}{c}}
{\!\!\!\!\text{if }{\Gamma _k^{\rm{a}}} \ge T_h^{\rm{a}},\!\!}&{\delta _0 > \delta _1}
\end{array}}\\
{\begin{array}{*{20}{c}}
{\!\!\!\!\text{if }{\Gamma _k^{\rm{a}}} < T_h^{\rm{a}},\!\!}&{\delta _0 \le \delta _1}
\end{array}}
\end{array}} \right.\\
\hat d\left[ k \right] = 1,\left\{ {\begin{array}{*{20}{l}}
{\begin{array}{*{20}{c}}
{\!\!\!\!\text{if }{\Gamma _k^{\rm{a}}} < T_h^{\rm{a}},\!\!}&{\delta _0 > \delta _1}
\end{array}}\\
{\begin{array}{*{20}{c}}
{\!\!\!\!\text{if }{\Gamma _k^{\rm{a}}} \ge T_h^{\rm{a}},\!\!}&{\delta _0 \le \delta _1}
\end{array}}
\end{array}} \right.
\end{array}
\end{align}
where $\Gamma _k^{\rm{a}} = \frac{1}{N}\sum\nolimits_{n = (k - 1)N + 1}^{kN} {{{\left| {{y^{\rm{a}}}\left[ n \right]} \right|}^2}} $ is the total energy of the $N$ consecutive samples corresponding to ${d\left[ k \right]}$, ${\delta _i} = \mathbb{E}\left[ {{\Gamma _k^{\rm{a}}}|{\mathcal{H}_i}} \right]$, $T_h^{\rm{a}}$ is the symbol detection threshold, and $\hat d\left[ k \right]$ is the estimated value of ${d\left[ k \right]}$.
\subsection{BER Performance with LNA Integration}
{\color{black}Accordingly to the central limit theorem{\footnote{\color{black}{The central limit theorem requires independent and identically distributed random variables and a large $N$, e.g., $N \ge 50$. It should be noted that $N \ge 50$ is practical in AmBC and the reason can be referred to the tenth footnote \cite{11263911}.}}}, $\Gamma_k^{\rm{a}}$ can be approximated as a Gaussian random variable with mean ${\delta _i}$ and variance $\varsigma _i^2$ under hypothesis ${\mathcal{H}_i}$.} Thus, the PDF of $\Gamma_k^{\rm{a}}$ under hypothesis ${\mathcal{H}_i}$ is approximately given by
\begin{align}
{\tilde f_{\Gamma _k^{\rm{a}}}}\left( {x|{{\cal H}_i}} \right) = \frac{1}{{\sqrt {2\pi {\varsigma _i}^2} }}\exp \left[ { - \frac{{{{\left( {x - {\delta _i}} \right)}^2}}}{{2\varsigma _i^2}}} \right].
\end{align}

\textbf{Theorem 1.} The BER for the receiver equipped with an LNA is derived as
\begin{align}\label{21}
\mathbb{P}_b^{\rm{a}} = \frac{1}{2}Q\left( {\frac{{T_h^{\rm{a}} - {\delta _{\min }}}}{{\sqrt {\varsigma _{\min }^2} }}} \right) + \frac{1}{2}Q\left( {\frac{{{\delta _{\max }} - T_h^{\rm{a}}}}{{\sqrt {\varsigma _{\max }^2} }}} \right),
\end{align}
where ${\delta _{\min }} = \min \left\{ {{\delta _0},{\delta _1}} \right\}$, ${\delta _{\max }} = \max \left\{ {{\delta _0},{\delta _1}} \right\}$, $\varsigma _{\min }^2 = \min \left\{ {\varsigma _0^2,\varsigma _1^2} \right\}$, $\varsigma _{\max }^2 = \max \left\{ {\varsigma _0^2,\varsigma _1^2} \right\}$.

\emph{Proof.} If ${\delta _0} < {\delta _1}$, the BER can be found as
\begin{align}\label{24}
{\mathbb{P}_b^{\rm{a}}}&= \Pr \left( {{\mathcal{H}_1}} \right)\Pr \left( {{\Gamma _k^{\rm{a}}} < T_h^{\rm{a}}|{\mathcal{H}_1}} \right) + \Pr \left( {{\mathcal{H}_0}} \right)\Pr \left( {{\Gamma _k^{\rm{a}}} > T_h^{{\rm{a}}}|{\mathcal{H}_0}} \right)\notag\\
 &= \frac{1}{2}\int_{ - \infty }^{T_h^{\rm{a}}} {{{\tilde f}_{{\Gamma _k^{\rm{a}}}}}\left( {x|{\mathcal{H}_1}} \right)} dx + \frac{1}{2}\int_{T_h^{\rm{a}}}^{ \infty } {{{\tilde f}_{{\Gamma _k^{\rm{a}}}}}\left( {x|{\mathcal{H}_0}} \right)} dx\notag\\
  &= {\frac{1}{2}Q\left( {\frac{{{\delta _1} - T_h^{\rm{a}}}}{{\sqrt {\varsigma _1^2} }}} \right) + \frac{1}{2}Q\left( {\frac{{T_h^{\rm{a}} - {\delta _0}}}{{\sqrt {\varsigma _0^2} }}} \right)}.
\end{align}
If ${\delta _0} > {\delta _1}$, the BER is similarly derived as
\begin{align}\label{25}
\mathbb{P}_b^{\rm{a}} = \frac{1}{2}Q\left( {\frac{{T_h^{\rm{a}} - {\delta _1}}}{{\sqrt {\varsigma _1^2} }}} \right) + \frac{1}{2}Q\left( {\frac{{{\delta _0} - T_h^{\rm{a}}}}{{\sqrt {\varsigma _0^2} }}} \right).
\end{align}
 Based on (\ref{24}) and (\ref{25}), we have (\ref{21}).
\hfill {$\blacksquare $}

It is clear that $\mathbb{P}_b^{\rm{a}}$ is a function of the expectation and variance of $\Gamma_k^{\rm{a}}$. This motivates us to derive these values, as summarized in \textbf{Theorem 2}.

\textbf{Theorem 2.} For given ${\beta _1}$ and ${\beta _3}$, the mean and variance of ${\Gamma _k^{\rm{a}}}$ are, respectively, calculated as (\ref{5}) and (\ref{6}), as shown at the top of next page,
\begin{figure*}[t]
\begin{align}\label{5}
\begin{array}{*{20}{c}}
{\mathbb{E}\left[ {{\Gamma _k^{\rm{a}}}\left| {{\mathcal{H}_i}} \right.} \right] = \beta _1^2{P_i} + 6\beta _3^2P_i^3 + 4{\beta _1}{\beta _3}P_i^2 + {N_{\rm{aw}}},}&{i = 0,1},
\end{array}
\end{align}
\hrulefill
\end{figure*}
\begin{figure*}[t]
\begin{align}\label{6}
\begin{array}{*{20}{c}}
{{{\mathop{\rm var}} \left[ {\Gamma _{_k}^{\rm{a}}\left| {{{\cal H}_i}} \right.} \right]} = \frac{1}{N}\left\{ \begin{array}{l}
\beta _1^4P_i^2 + 16\beta _1^3{\beta _3}P_i^3 + 116\beta _1^2\beta _3^2P_i^4 + 432{\beta _1}\beta _3^3P_i^5 + 684\beta _3^4P_i^6\\
 + 2\beta _1^2{P_i}{N_{\rm{aw}}} + 8{\beta _1}{\beta _3}P_i^2{N_{\rm{aw}}} + 12\beta _3^2P_i^3{N_{\rm{aw}}} + N_{\rm{aw}}^2
\end{array} \right\},}&{i = 0,1}
\end{array}
\end{align}
\hrulefill
\end{figure*}
where ${P_i} = {\left| {{h_i}} \right|^2}{P_s}$.

\emph{Proof.} Based on (\ref{2}), ${\left| {y_k^{\rm{a}}\left[ n \right]} \right|^2}$ can be calculated as
\begin{align}
{\left| {y_k^{\rm{a}}\left[ n \right]} \right|^2} &= {\left| {x\left[ n \right] + w_{\rm{a}}\left[ n \right]} \right|^2}\notag \\
 &= {\left| {x\left[ n \right]} \right|^2} + {\left| {w_{\rm{a}}\left[ n \right]} \right|^2} + 2{\rm{Re}}\left[ {x\left[ n \right]w_{\rm{a}}^*\left[ n \right]} \right],
\end{align}
where $x\left[ n \right] = {h_i}s\left[ n \right]\left( {{\beta _1} + {\beta _3}{Z_i}} \right)$, $Z_i = {\left| {{h_i}s\left[ n \right]} \right|^2}$. Due to $Z_i \sim \exp \left( {\frac{1}{{{P_i}}}} \right)$, based on the $m$-th order moment formula of the exponential distribution we have $\mathbb{E}\left[ {{Z_i^m}} \right] = m!P_i^m$.

{It can be derived that $\mathbb{E}\left\{ {{\mathop{\rm Re}\nolimits} \left[ {x\left[ n \right]{w_{\rm{a}}^*}\left[ n \right]} \right]} \right\} = 0$, which further gives the following results, i.e., $\mathbb{E}\left[ {{{\left| {y_k^{\rm{a}}\left[ n \right]} \right|}^2}} \right] = \mathbb{E}\left[ {{{\left| {x\left[ n \right]} \right|}^2}} \right] + \mathbb{E}\left[ {{{\left| {w_{\rm{a}}\left[ n \right]} \right|}^2}} \right]$, ${\mathbb{E}\left[ {{{\left| {x\left[ n \right]} \right|}^2}} \right] = \mathbb{E}\left[ {{Z_i}{{\left( {{\beta _1} + {\beta _3}{Z_i}} \right)}^2}} \right]}= \beta _1^2{P_i} + 6\beta _3^2P_i^3 + 4{\beta _1}{\beta _3}P_i^2$, and $ \mathbb{E}\left[ {{{\Gamma _k^{\rm{a}}}\left| {{\mathcal{H}_i}} \right.}} \right] = \frac{1}{N}\sum\limits_{n = (k - 1)N + 1}^{kN} \mathbb{E} \left\{ {{{\left| {y_k^{\rm{a}}\left[ n \right]} \right|}^2}} \right\}=
{\beta _1^2{P_i} + 6\beta _3^2P_i^3 + 4{\beta _1}{\beta _3}P_i^2 + {N_{\rm{aw}}},} {i = 0,1}$.

The variance of ${\Gamma _k^{\rm{a}}}$ is written as  ${\rm{var}}\left[ {\Gamma _k^{\rm{a}}\left| {{H_i}} \right.} \right] = \frac{1}{{{N^2}}}\sum\limits_{n = (k - 1)N + 1}^{kN} {{\rm{var}}\left\{ {{{\left| {y_k^{\rm{a}}\left[ n \right]} \right|}^2}} \right\}}  = \frac{1}{{{N^2}}}\sum\limits_{n = (k - 1)N + 1}^{kN} {\left[ {E\left\{ {{{\left| {y_k^{\rm{a}}\left[ n \right]} \right|}^4}} \right\} - {E^2}\left\{ {{{\left| {y_k^{\rm{a}}\left[ n \right]} \right|}^2}} \right\}} \right]} $. 

Based on (\ref{2}), ${{{\left| {y_k^{\rm{a}}\left[ n \right]} \right|}^4}}$ can be calculated as (\ref{15}), as shown at the top of next page.
\begin{figure*}[t]
\begin{align}\label{15}
{\left| {y_k^{\rm{a}}\left[ n \right]} \right|^4} &= {\left| {x\left[ n \right]} \right|^4} + {\left| {w_{\rm{a}}\left[ n \right]} \right|^4} + {\left( {x\left[ n \right]{w_{\rm{a}}^*}\left[ n \right]} \right)^2} + {\left( {{x^*}\left[ n \right]w_{\rm{a}}\left[ n \right]} \right)^2} + 2{\left| {x\left[ n \right]} \right|^2}{\left| {w_{\rm{a}}\left[ n \right]} \right|^2} + 2{\left| {x\left[ n \right]} \right|^2}x\left[ n \right]{w_{\rm{a}}^*}\left[ n \right]\notag\\
 &+ 2{\left| {x\left[ n \right]} \right|^2}{x^*}\left[ n \right]w_{\rm{a}}\left[ n \right] + 2{\left| {w_{\rm{a}}\left[ n \right]} \right|^2}x\left[ n \right]{w_{\rm{a}}^*}\left[ n \right] + 2{\left| {w_{\rm{a}}\left[ n \right]} \right|^2}{x^*}\left[ n \right]w_{\rm{a}}\left[ n \right] + 2x\left[ n \right]{w_{\rm{a}}^*}\left[ n \right]{x^*}\left[ n \right]w_{\rm{a}}\left[ n \right]
\end{align}
\hrulefill
\end{figure*}
Since $\mathbb{E}\left\{ {{{\left( {x\left[ n \right]{w_{\rm{a}}^*}\left[ n \right]} \right)}^2}} \right\} = \mathbb{E}\left\{ {{{\left( {{x^*}\left[ n \right]w_{\rm{a}}\left[ n \right]} \right)}^2}} \right\} = 0$ holds, we have
$\mathbb{E}\left[ {{{\left| {y_k^{\rm{a}}\left[ n \right]} \right|}^4}} \right] = \mathbb{E}\left[ {{{\left| {x\left[ n \right]} \right|}^4}} \right] + \mathbb{E}\left[ {{{\left| {w_{\rm{a}}\left[ n \right]} \right|}^4}} \right]
 + 4\mathbb{E}\left[ {{{\left| {x\left[ n \right]} \right|}^2}} \right]\mathbb{E}\left[ {{{\left| {w_{\rm{a}}\left[ n \right]} \right|}^2}} \right]$,
where $\mathbb{E}\left[ {{{\left| {w_{\rm{a}}\left[ n \right]} \right|}^2}} \right] = {N_{\rm{aw}}}$, $\mathbb{E}\left[ {{{\left| {w_{\rm{a}}\left[ n \right]} \right|}^4}} \right] = 2{N_{\rm{aw}}}$, and
$\mathbb{E}\left[ {{{\left| {x\left[ n \right]} \right|}^4}} \right]
= \mathbb{E}\left[ {{Z_i^2}{{\left( {{\beta _1} + {\beta _3}Z_i} \right)}^4}} \right]
= \sum\limits_{m = 0}^4 {\left( {\begin{array}{*{20}{c}}
4\\
m
\end{array}} \right)} \beta _1^{4 - m}\beta _3^m\mathbb{E}\left[ {{Z_i^{m + 2}}} \right]
 = 2\beta _1^4P_i^2 + 24\beta _1^3{\beta _3}P_i^3 + 144\beta _1^2\beta _3^2P_i^4 + 480{\beta _1}\beta _3^3P_i^5.$


By substituting $\mathbb{E}\left[ {{{\left| {x\left[ n \right]} \right|}^2}} \right]$ and $\mathbb{E}\left[ {{{\left| {x\left[ n \right]} \right|}^4}} \right]$ into $\mathbb{E}\left[ {{{\left| {y_k^{\rm{a}}\left[ n \right]} \right|}^4}} \right]$, then substituting $ \mathbb{E}\left[ {{{\Gamma _k^{\rm{a}}}\left| {{\mathcal{H}_i}} \right.}} \right]$ and the result obtained from $\mathbb{E}\left[ {{{\left| {y_k^{\rm{a}}\left[ n \right]} \right|}^4}} \right]$ into ${\rm{var}}\left[ {\Gamma _k^{\rm{a}}\left| {{{\cal H}_i}} \right.} \right]$, we obtain (\ref{6}).}
\hfill {$\blacksquare $}

 Using the mean and variance of ${\Gamma _k^{\rm{a}}}$ provided in \textbf{Theorem 2}, we derive the near-optimal detection threshold to minimize the BER. The result is summarized in \textbf{Theorem 3}.

\textbf{Theorem 3.} The expressions of the near-optimal detection threshold with LNA is given by
\begin{align}\label{22}
T_{h,opt}^{\rm{a}} = \frac{{\bar \delta  + \sqrt {{C_\varsigma }{{({\delta _0} - {\delta _1})}^2} + {C_\varsigma }(\varsigma _1^2 - \varsigma _0^2)\ln {C_\varsigma }} }}{{{C_\varsigma } - 1}},
\end{align}
where ${C_\varsigma } = \frac{{\varsigma _1^2}}{{\varsigma _0^2}}$, $\bar \delta  = {\delta _0}{C_\varsigma } - {\delta _1}$.

\emph{Proof.} 
The threshold ${{\Gamma _k^{\rm{a}}}}$ can be computed from
\begin{align}\label{26}
{\tilde f_{{\Gamma _k^{\rm{a}}}}}\left( {T_h^{\rm{a}}|{\mathcal{H}_0}} \right) = {\tilde f_{{\Gamma _k^{\rm{a}}}}}\left( {T_h^{\rm{a}}|{\mathcal{H}_1}} \right).
\end{align}
Taking the natural logarithm of both sides of (\ref{26}) and solving the resulting system of equations, we obtain (\ref{22}).\hfill {$\blacksquare $}
\subsection{Theoretical Analysis of LNA-Induced Performance Gain}
The DC serves as a valuable metric for the symbol detection performance since it characterizes the variance-normalized distance between the centers of two PDFs \cite{395235}. While no direct relationship exists between the DC and the BER, a higher DC is a reliable indicator of better BER performance. Therefore, comparing the DCs clearly demonstrates the contribution of the LNA to enhancing the symbol detection performance.

For the traditional AmBC receiver without LNA, the DC is given by
\begin{align}\label{27}
d_{\rm{noLNA}}^2 = \frac{{{{\left( {\mathbb{E}[{\Gamma _k}\left| {{\mathcal{H}_1}} \right.] - \mathbb{E}[{\Gamma _k}\left| {{\mathcal{H}_0}} \right.]} \right)}^2}}}{{{\mathop{\rm var}} [{\Gamma _k}\left| {{\mathcal{H}_0}} \right.]}} = N{\left( {\frac{{{P_1} - {P_0}}}{{{P_0} + {N_w}}}} \right)^2},
\end{align}
where, ${N_w} = {N_{\rm{ar}}} + {N_{{\rm{cov}}}} + {\alpha ^2}{\left| {{h_{tr}}} \right|^2}{N_{\rm{at}}}$, $\mathbb{E}[{\Gamma _k}\left| {{\mathcal{H}_i}} \right.] = {P_i} + {N_w}$, ${\mathop{\rm var}} [{\Gamma _k}\left| {{\mathcal{H}_i}} \right.] = \frac{1}{N}\left[ {P_i^2 + 2{P_i}{N_w} + N_w^2} \right]$ (see eqs. (22) in \cite{8007328}).

The DC with LNA as follows,
\begin{align}\label{23}
d_{{\rm{LNA}}}^2 \simeq N\beta _1^4{\left( {\frac{{{P_1} - {P_0}}}{{\beta _1^2{P_0} + {N_{{\rm{aw}}}}}}} \right)^2} = N{\left( {\frac{{{P_1} - {P_0}}}{{{P_0} + \frac{{{N_{{\rm{aw}}}}}}{{\beta _1^2}}}}} \right)^2},
\end{align}
where $\frac{{{N_{\rm{aw}}}}}{{\beta _1^2}} = {N_{\rm{ar}}} + \frac{{{N_{{\rm{cov}}}}}}{{\beta _1^2}} + {\alpha ^2}{\left| {{h_{tr}}} \right|^2}{N_{\rm{at}}}$.

\emph{Proof. }Please refer to Appendix.
\hfill {$\blacksquare $}

\textit{Remark 1. } It is evident from (\ref{27}) and (\ref{23}) that $d_{\rm{LNA}}^2 > d_{\rm{noLNA}}^2$ holds due to ${N_w} > \frac{{{N_{{\rm{aw}}}}}}{{\beta _1^2}}$. We also note that ${P_0}$ dominates the denominators of both DCs if the transmit power of the ambient source is sufficiently large. In this case, the two coefficients converge to $\frac{{N{{\left( {{P_1} - {P_0}} \right)}^2}}}{{P_0^2}}$, i.e., $d_{{\rm{LNA}}}^2 \simeq d_{{\rm{noLNA}}}^2$, due to the negligibility  of ${{N_{{\rm{aw}}}}}$ and ${\frac{{{N_{{\rm{aw}}}}}}{{\beta _1^2}}}$ compared to ${P_0}$. {\color{black} The above observations indicate that the advantage of the LNA lies in the low-to-moderate ambient source transmission power region, but diminishes at high transmit power levels.}



\subsection{Threshold Required Parameters Estimation}
 For the proposed near-optimal detection threshold (\ref{22}), we should  estimate the parameters ${\delta _i}$ and ${\varsigma _i}^2$. Here, we estimate ${\delta _i}$ and ${\varsigma _i}^2$ by utilizing the energy of tag's pilot symbols, and subsequently substitute them into (\ref{22}) to obtain the estimated threshold $\hat T_{h,opt}^{\rm{a}}$. The estimation method is summarized in Algorithm 1.
\begin{table}[!ht]
\centering
\begin{tabular}{l}
  \toprule
  \textbf{Algorithm 1} The parameter estimation method with LNA \\
  \midrule
  \textbf{Input:} Received samples $y_k^{\rm{a}}(n)$ for pilot symbols, pilot symbols \\
  length $K_{\mathrm{train}}$, $N$\\
  \textbf{Output:} The parameter ${\hat \delta _0}$, ${\hat \delta _1}$, $\hat \varsigma _0^2$ and $\hat \varsigma _1^2$\\
  \: 1: \textbf{Step 1: Energy Calculation} \\
  \: 2: For each pilot symbol $k = 1, 2, \ldots, K_{\mathrm{train}}$, compute: \\
  \: 3: \quad ${A_k} = \frac{1}{N}\sum\limits_{n = \left( {k - 1} \right)N + 1}^{kN} {{{\left| {y_k^{{\rm{a}}}\left( n \right)} \right|}^2}}$\\
  \: 4: \textbf{Step 2: Direct Grouping} \\
  \: 5: Separate energy values into two groups: \\
  \: 6: \quad Group 0: $\{A_k \mid d[k] = 0\}$ \\
  \: 7: \quad Group 1: $\{A_k \mid d[k] = 1\}$ \\
  \: 8: \textbf{Step 3: Parameter Estimation} \\
  \: 9: Compute sample mean and variance for each group: \\
  \: 10: \quad ${\hat \delta _0} = {\hat E_0} = \frac{2}{{{K_{{\rm{train}}}}}}\sum\limits_{{A_k} \in {\rm{Group}}\;{\rm{0}}} {{A_k}} $,\\
  \: 11:  \quad $\hat \varsigma _0^2 = {\hat D_0} = \left( {\frac{2}{{{K_{{\rm{train}}}} - 2}}} \right)\sum\limits_{{A_k} \in {\rm{Group}}\;{\rm{0}}} {{{({A_k} - {{\hat E}_0})}^2}} $ \\
  \: 12:  \quad  ${\hat \delta _1} = {\hat E_1} = \frac{2}{{{K_{{\rm{train}}}}}}\sum\limits_{{A_k} \in {\rm{Group}}\;{\rm{1}}} {{A_k}} $,\\
  \: 13: \quad $\hat \varsigma _1^2 = {\hat D_1} = \frac{2}{{{K_{{\rm{train}}}} - 2}}\sum\limits_{{A_k} \in {\rm{Group}}\;{\rm{1}}} {{{({A_k} - {{\hat E}_1})}^2}} $ \\
  \: 14: \textbf{return} ${\hat \delta _0}$, ${\hat \delta _1}$, $\hat \varsigma _0^2$ and $\hat \varsigma _1^2$\\
  \bottomrule
\end{tabular}
\end{table}\vspace{-10pt}
\section{Simulation Results}

\begin{figure*}[htbp]
    \centering
    \begin{minipage}{0.64\linewidth} 
        \centering
        \begin{subfigure}[b]{0.5\linewidth}
            \centering
            \includegraphics[width=\linewidth]{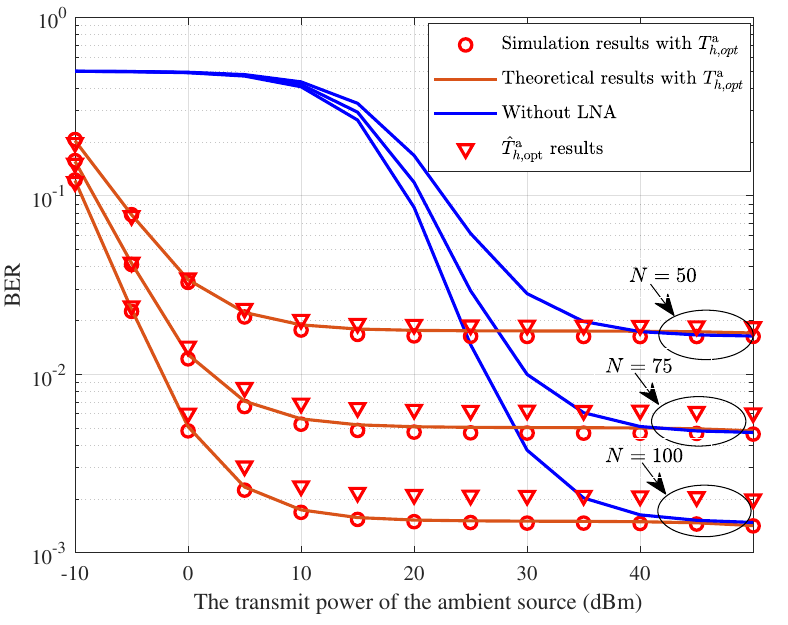}
            \caption{\small{BER versus the transmit power of the ambient source.}}
            \label{chutian2}
        \end{subfigure}
        \hfill
        \begin{subfigure}[b]{0.49\linewidth}
            \centering
            \includegraphics[width=\linewidth]{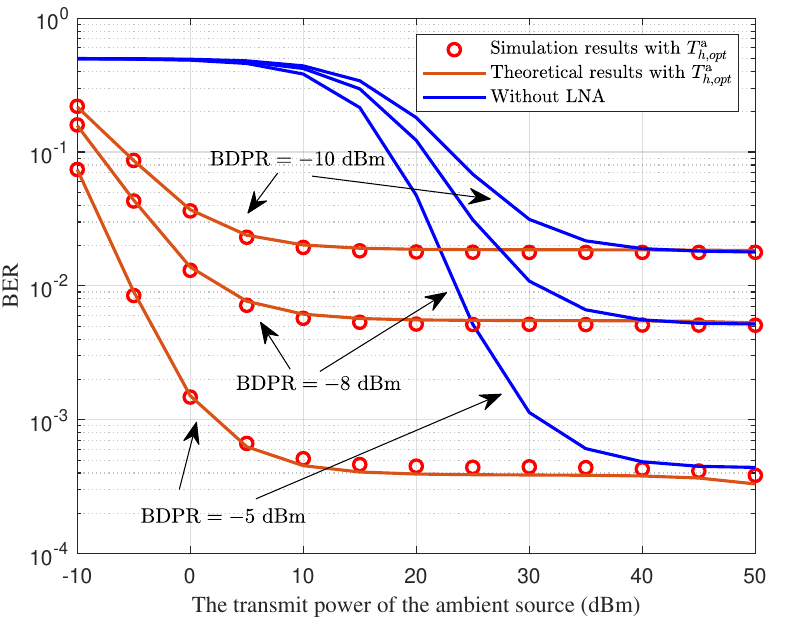}
            \caption{\small{BER versus the transmit power of the ambient source under different BDPR.}}
            \label{chutian3}
        \end{subfigure}
\vspace{-0.5em}
       \caption{\small BER performance under different ambient source power levels and link quality conditions.}
    \end{minipage}
    \hfill
     \vspace{-0.8em}
    \begin{minipage}{0.34\linewidth} 
        \centering
        \includegraphics[width=\linewidth]{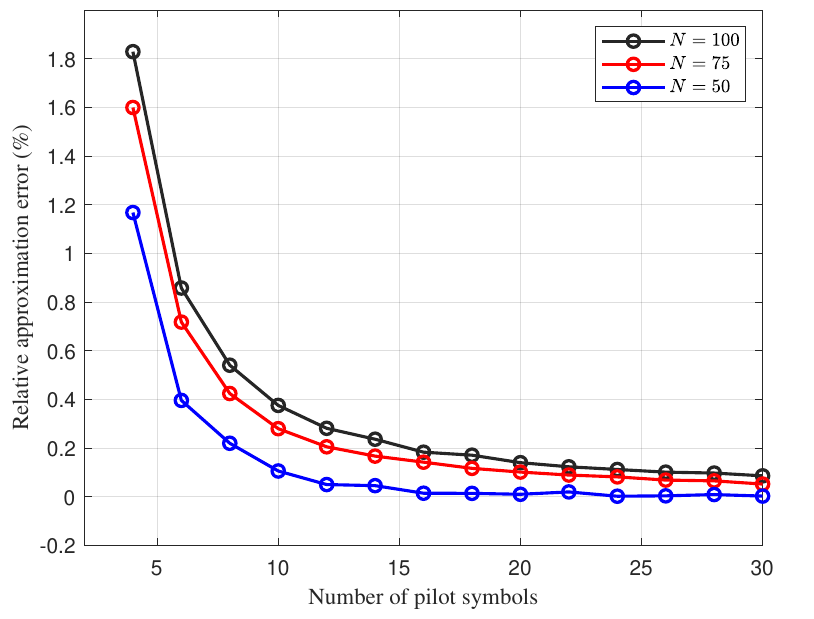}
        \caption{\small{Relative approximation error versus number of pilot symbols.}}
        \label{fig5}
    \end{minipage}

\end{figure*}
\begin{figure*}[t]
\begin{align}\label{30}
d_{\rm{LNA}}^2 &= \frac{{{{\left( {\mathbb{E}[\Gamma _k^a\left| {{H_1}} \right.] - \mathbb{E}[\Gamma _k^a\left| {{H_0}} \right.]} \right)}^2}}}{{{\mathop{\rm var}} [{\Gamma _k}\left| {{H_0}} \right.]}}
 = N\frac{{{{\left[ {\beta _1^2\left( {{P_1} - {P_0}} \right) + 6\beta _3^2\left( {P_1^3 - P_0^3} \right) + 4{\beta _1}{\beta _3}\left( {P_1^2 - P_0^2} \right)} \right]}^2}}}{{\left( \begin{array}{l}
\beta _1^4P_0^2 + 16\beta _1^3{\beta _3}P_0^3 + 116\beta _1^2\beta _3^2P_0^4 + 432{\beta _1}\beta _3^3P_0^5 + 684\beta _3^4P_0^6\\
 + 2\beta _1^2{P_0}{N_{aw}} + 8{\beta _1}{\beta _3}P_0^2{N_{aw}} + 12\beta _3^2P_0^3{N_{aw}} + N_{aw}^2
\end{array} \right)}}
\end{align}
\hrulefill
\end{figure*}

In this section, we resort to numerical examples to validate our derived results and analysis. \textcolor{black}{We amuse that all channels undergo independent quasi-static Rayleigh fading. Let ${h_0} \sim \mathbb{CN}(0,r_0^{ - {v_0}})$, ${h_{st}} \sim \mathbb{CN}(0,r_{st}^{ - {v_{st}}})$ and ${h_{tr}} \sim \mathbb{CN}(0,r_{tr}^{ - {v_{tr}}})$, where ${v_0}$, ${v_{st}}$, and ${v_{tr}}$ are the path loss exponents, and $r_0$, $r_{st}$ and $r_{tr}$ represent the corresponding link distances.}
The parameters are set as follows: ${v_0} = 4.5$, ${v_{st}} = 4.5$, ${v_{tr}} = 2.5$ {\color{black}\cite{Gordon2011Principles}}, ${r_0} = 50$ m, ${r_{st}} = 50$ m, ${r_{tr}} = 10$ m, ${w_{\rm{ar}}} = {w_{\rm{at}}} =  - 100$ dBm, ${w_{{\mathop{\rm cov}} }} =  - 70$ dBm, $K = 100$, and $\alpha  =  - 1.1$ dB {\color{black}\cite{kellogg2016passive}}.

Fig. 4 (a) plots the BER versus the transmit power of the source ${P_s}$ with LNA integration. The threshold is calculated by (\ref{22}). It can be seen that our derived BER matches simulation results well, and the accuracy of (\ref{22}) improves as $N$ increases. As the ${P_s}$ increases, the BER decreases.  {\color{black}For comparative analysis, a scenario without LNA is provided, showing that the receiver equipped with an LNA exhibits significantly better performance compared to the case without LNA at low-to-medium ${P_s}$.} Nevertheless, as the ${P_s}$ increases further, the advantage of the LNA diminishes, as expected in \textit{Remark 1}. {\color{black}Additionally, it is evident that after the application of LNA, there is a noticeable error floor when the ${P_s}$ is 20 dBm. To investigate whether this state transition point is influenced by the power ratio between the backscatter link and the direct link (BDPR), we present Fig. 4 (b), which illustrates the BER versus the ${P_s}$ under different BDPR conditions, with \( N \) set to 75. The results indicate that when the ${P_s}$ is in the low-to-medium range, the configuration of LNA significantly enhances detection performance regardless of BDPR, and the power starting point of the error floor remains unchanged. This demonstrates that relative BDPR only affects the degree of performance improvement without altering the applicability of LNA. In summary, the LNA can significantly improve the AmBC symbol detection performance at low-to-moderate ${P_s}$.}



Fig. 5 illustrates the variation of the relative approximation error $R$ between the theoretical and estimated detection thresholds with respect to the number of pilot symbols. Following \cite{8633928}, $R = \frac{{\left| {T_{h,opt}^{\rm{a}} - \hat T_{h,opt}^{\rm{a}}} \right|}}{{\hat T_{h,opt}^{\rm{a}}}}$. It can be observed that the estimated detection threshold exhibits high accuracy, with the error decreasing as the number of pilot symbols increases. This reduction is particularly pronounced in the low pilot region (less than $10\% $ of the total symbols), and after exceeding $20\% $ pilot overhead, additional pilots do not significantly improve accuracy. {\color{black}In Fig. 4 (a), the pilot overhead is set to 20\%, and the results indicate that the BER curve based on $\hat T_{h,opt}^{\rm{a}}$ closely matches the theoretically near-optimal curve.}
\section{Conclusions}
In this paper, we investigated the symbol detection performance of an AmBC receiver equipped with an LNA.  {\color{black}We proposed a new detection framework that incorporates the effect of the LNA, derived the BER of ED with the near-optimal detection threshold, and demonstrated the advantage of using LNA at low-to-moderate ambient source transmission power.} We also proposed a practical method to estimate the parameters required by calculating the near-optimal detection threshold by exploiting the tag's pilot symbols. {\color{black}In future work, we will investigate the joint design of LNAs and various advanced detection schemes.}
\appendix
Based on (\ref{5}) and (\ref{6}), the DC after applying the LNA can be expressed as (\ref{30}), as shown at the top of this page.
Since ${P_i}$ is extremely small, typically on the order of ${ - 50}$dBm or even smaller, higher-order terms such as ${P_i^3}$ and beyond can be considered negligible in comparison. Therefore, by omitting these higher-order terms, the approximate DC with the LNA can be expressed as (\ref{23}).
\bibliographystyle{IEEEtran}
\bibliography{ref}

\end{document}